\documentclass[%
 prl,
 jmp,%
 amsmath,amssymb,
 floatfix,
reprint,%
 showpacs,%
]{revtex4-1}

\usepackage{comment}
\usepackage{amssymb}
\usepackage{amsmath}
\usepackage{mathtools}
\usepackage{relsize}
\usepackage{array}
\usepackage{bm}
\usepackage{graphicx}
\usepackage{bigints}
\usepackage{ifthen}
\usepackage{multirow}
\usepackage{threeparttable}
\usepackage{rotating}
\usepackage{hyperref}
\usepackage{xcolor}
\usepackage{calrsfs}
\usepackage{empheq}
\usepackage[normalem]{ulem}
\usepackage{media9}
\usepackage[margin=0pt,font=small,labelfont=normalfont,labelsep=period,format=plain,justification=centerlast]{caption}
\usepackage{xargs}   

\usepackage[draft,footnote,marginclue,innerlayout=inline]{fixme}
\fxloadlayouts{marginnote}
\fxsetup{theme=color,mode=multiuser}
\FXRegisterAuthor{V}{anV}{{\color{red}{VG}}}

\DeclareGraphicsExtensions{.pdf,.png,.eps}
\graphicspath{{figs/}{./}}
\usepackage{cleveref}
\crefname{equation}{}{}
\Crefname{equation}{}{}
\crefname{figure}{Fig.~\!\!\!}{Figs.~\!\!\!}
\Crefname{figure}{Fig.~\!\!\!}{Figs.~\!\!\!}

\newcommand{\paren}[1]{\left( #1 \right)}

\usepackage[colorinlistoftodos,prependcaption,textsize=tiny]{todonotes}
\newcommandx{\lrfnote}[2][1=]{\todo[linecolor=purple,backgroundcolor=purple!25,bordercolor=purple,#1]{#2}}
\newcommandx{\trnote}[2][1=]{\todo[linecolor=red,backgroundcolor=red!25,bordercolor=red,#1]{TR: #2}}
\newcommandx{\dbnote}[2][1=]{\todo[linecolor=blue,backgroundcolor=blue!25,bordercolor=blue,#1]{#2}}
\newcommandx{\lwnote}[2][1=]{\todo[linecolor=olive,backgroundcolor=olive!25,bordercolor=olive,#1]{#2}}
\newcommandx{\thiswillnotshow}[2][1=]{\todo[disable,#1]{#2}}

\usepackage{tikz}
\usetikzlibrary{intersections,arrows,decorations,decorations.pathmorphing,calc}
\usepackage{pgfplots}
\pgfplotsset{compat=newest}
\pgfplotsset{plot coordinates/math parser=false}

\newcommand{\der}[2]{\frac{d {#1}}{d {#2}}}
\newcommand{\derp}[2]{\frac{\partial {#1}}{\partial {#2}}}

\makeatletter
\let\jnl@style=\rmfamily
\def\ref@jnl#1{{\jnl@style#1}}
\@ifundefined{aap}{\newcommand\aap{\ref@jnl{A\&A}}}{}%
\@ifundefined{apj}{\newcommand\apj{\ref@jnl{ApJ}}}{}%
\@ifundefined{apjl}{\newcommand\apjl{\ref@jnl{Astrophys.~J.~Lett.}}}{}%
\@ifundefined{ssr}{\newcommand\ssr{\ref@jnl{Space~Sci.~Rev.}}}{}%
\@ifundefined{jgr}{\newcommand\jgr{\ref@jnl{J.~Geophys.~Res.}}}{}%
\@ifundefined{jatp}{\newcommand\jatp{\ref@jnl{J.~Atmos.~Terr.~Phys.}}}{}%
\@ifundefined{solphys}{\newcommand\solphys{\ref@jnl{Sol.~Phys.}}}{}%
\@ifundefined{pp}{\newcommand\pp{\ref@jnl{Phys.~Plasmas}}}{}%
\@ifundefined{mnras}{\newcommand\mnras{\ref@jnl{MNRAS}}}{}%
\@ifundefined{apjs}{\newcommand\apjs{\ref@jnl{ApJS}}}{}%
\@ifundefined{azh}{\newcommand\azh{\ref@jnl{AZh}}}{}%
\catcode`\@=11

\def\beq#1\eeq{\begin{equation}#1\end{equation}}
\def\bes#1\ees{\begin{subequations}#1\end{subequations}}
\def\bea#1\eea{\begin{align}#1\end{align}}
\def\n{\nonumber\\}

\def\o{\omega}

\usepackage{scalerel}

\allowdisplaybreaks

\def\A{\tilde{A}}
\def\t{\tau}
\def\o{\tilde{\omega}}
\def\k{\tilde{k}}
\let\PP\P
\def\P{{P}}

\begin{document}

\title{Neuronal avalanches and critical dynamics of brain waves}

\author{Vitaly L. Galinsky}
\email{vit@ucsd.edu}
\affiliation{Center for Scientific Computation in Imaging,
University of California at San Diego, La Jolla, CA 92037-0854, USA}
\author{Lawrence R. Frank}
\email{lfrank@ucsd.edu}
\affiliation{Center for Scientific Computation in Imaging,
University of California at San Diego, La Jolla, CA 92037-0854, USA}
\affiliation{
Center for Functional MRI,
University of California at San Diego, La Jolla, CA 92037-0677, USA}

\date{\today}

\begin{abstract}

Analytical expressions for scaling of brain wave spectra derived from
the general nonlinear wave Hamiltonian form show excellent agreement
with experimental ''neuronal avalanche'' data.  The theory of the
weakly evanescent nonlinear brain wave dynamics
\citep{Galinsky:2020a,*Galinsky:2020b} reveals the underlying
collective processes hidden behind the phenomenological statistical
description of the neuronal avalanches and connects together the whole
range of brain activity states, from oscillatory wave-like modes, to
neuronal avalanches, to incoherent spiking, showing that the neuronal
avalanches are just the manifestation of the different nonlinear side
of wave processes abundant in cortical tissue.  In a more broad way
these results show that a system of wave modes interacting through all
possible combinations of the third order nonlinear terms described by
a general wave Hamiltonian necessarily produces anharmonic wave modes
with temporal and spatial scaling properties that follow scale free
power laws. To the best of our knowledge this was never reported in
the physical literature and may be applicable to many physical systems
that involve wave processes and not just to neuronal avalanches.
\end{abstract}


\maketitle

The standard view of brain electromagnetic
activity classifies this activity into two significant but
essentially independent classes.  The first class
includes a variety of the oscillatory and wave-like patterns that show
relatively high level of coherence across a wide range of spatial and
temporal scales \citep{buzsaki2006rhythms}.  The second class
focusses on the asynchronous, seemingly incoherent spiking activity
at scales of a single neuron and often uses various {\it ad hoc}
neuron models
\citep{pmid12991237,*pmid19431309,*Nagumo1962,*pmid7260316,*pmid18244602}
to describe this activity.  Linking these two seemingly
disparate classes to explain the emergence of oscillatory rhythms
from incoherent activity is essential to understanding brain
function and is typically posed in the form using the construct of
networks of incoherently spiking neurons
\citep{Gerstner:2014:NDS:2635959,*pmid33192427,*pmid33288909}.

Coherent macroscopic behavior arising from seemingly
incoherent microscopic processes naturally suggests the influence of
critical phenomena, a potential model from brain activity that was
bolstered by the experimental discovery of the ``neuronal
avalanches''\citep{pmid14657176,*pmid15175392} where both spatial and
temporal distributions of spontaneous propagating neuronal activity in
2D cortex slices were shown to follow scale-free power
laws. This discovery has generated significant interest in
the role and the importance of criticality in brain
activity
\citep{pmid23003192,*2010NatPh...6..744Cl,*pmid22701101,*pmid25009473,*pmid32503982,*pmid31172737},
especially for transmitting or processing information \citep{pmid33750159}.

Although the precise neuronal mechanisms leading to the observed
scale-free avalanche behavior is still uncertain after almost 20 years
since their discovery, the commonly agreed upon paradigm is
that this collective neuronal avalanche activity represents a
unique and specialized pattern of brain activity
that exists somewhere between the oscillatory, wave-like coherent
activity and the asynchronous and incoherent spiking.
Central to this claim of neuronal avalanches as a unique
brain phenomena is that they do not show either wave-like
propagation or synchrony at short scales, and thus
constitute a new mode of network activity
\citep{pmid14657176,*pmid15175392} that can be phenomenologically
described using the ideas of the self-organized criticality
\citep{PhysRevLett.59.381,*PhysRevA.38.364}, and extended to
the mean-field theory of the self-organized branching processes (SOBP)
\citep{PhysRevLett.75.4071,*PhysRevE.54.2483,*pmid12513377}.  

However, despite the success of the SOBP theory in
describing neuronal avalanche statistical properties, i.e.,
replicating the power law exponents based on the criticality
considerations, the SOBP theory provides no explanation about
the physical mechanisms of the critical behavior and its
relationship to the development of the observed collective neuronal
``avalanche'' behavior.  Because similar statistics can
result from several mechanisms other than 
critical dynamics \citep{PhysRevLett.97.118102,*pmid20161798,*pmid28208383},
it is essential to have a physical model that explains the
relationship between the statistical properties and the existence,
if any, of critical neural phenomena arising from the actual collective
behavior of neuronal populations.  While it is generally accepted in
that some form of critical phenomena is at
work, this has led to the presupposition of \textit{ad hoc}
descriptive models \citep{1997PhRvE..56..826R,*pmid28505725,*pmid12005890,*SantoE1356}
that exhibit critical behavior, but provide no
insight into the actual physical mechanisms that
might produce such critical dynamics.

In this Letter we show that our recently described theory of
weakly evanescent brain waves (WETCOW) originally developed
in \citep{Galinsky:2020a,*Galinsky:2020b} and then reformulated in a
general Hamiltonian framework \citep{Galinsky:2021a} provides
a physical theory, based on the propagation of
electromagnetic fields through the highly complex geometry of
inhomogeneous and anisotropic domain of real brain tissues, that explains
the broad range of observed seemingly disparate brain wave
characteristics.  This theory produces a set of nonlinear equations
for both the temporal and spatial evolution of
brain wave modes that includes all possible nonlinear interaction
between propagating modes at multiple spatial and temporal
scales and degrees of nonlinearity.  This theory bridges
the gap between the two seemingly unrelated spiking and wave 'camps' as the
generated wave dynamics includes the complete spectra of
brain activity ranging from incoherent asynchronous spatial or
temporal spiking events, to coherent wave-like propagating modes in
either temporal or spatial domains, to
collectively synchronized spiking of multiple temporal or spatial
modes. Consequently, we
demonstrate that the origin of these 'avalanche' properties emerges
directly from the same theory that produces this wide range of activity
and does not require one to posit the existence of either new brain
activity states, nor construct analogies between brain activity and
{\it ad hoc} generic 'sandpile' models.

Following \citep{Galinsky:2021a} we begin with a nonlinear
Hamiltonian form for an anharmonic wave mode 
\vspace*{-2pt}
\begin{align}
\label{eq:H}
H^s(a,a^\dag) &= 
\Gamma a a^\dag\! + a a^\dag\! \left[\beta_a a + \beta_{a^\dag} a^\dag\! -
  2\alpha \left(a a^\dag \right)^{1/2}\right] 
\end{align}
where $a$ is a complex wave amplitude and $a^\dag$ is its
conjugate. The amplitude $a$ denotes either temporal $a_k(t)$ 
or spatial $a_\omega(x)$ wave mode amplitudes
that are related to the spatiotemporal
wave field $\psi(x,t)$ through a Fourier integral expansions
\begin{align}
\label{eq:series}
a_k(t)=\frac{1}{2\pi} \int\limits_{-\infty}^{\infty}
\psi(x,t) e^{-i\left(k x + \omega_kt\right)}dx,\\
a_\omega(x) = \frac{1}{2\pi}\int\limits_{-\infty}^{\infty}
\psi(x,t) e^{-i\left(k_\omega x + \omega t\right)}dt,
\end{align}
where for the sake of clarity we use one dimensional scalar
expressions for spatial variables $x$ and $k$, but it can be easily
generalized for the multi dimensional wave propagation as well.  The
frequency $\omega$ and the wave number $k$ of the wave modes satisfy
dispersion relation $D(\omega,k)=0$, and $\omega_k$ and $k_\omega$
denote the frequency and the wave number roots of the dispersion
relation (the structure of the dispersion relation and its connection
to the brain tissue properties has been discussed in
\citep{Galinsky:2020a}). The multiple temporal $a_k(t)$
or spatial $a_\omega(x)$ wave mode amplitudes can be used to define
the time dependent wave number energy spectral density $\PP_k(t)$ or 
the position dependent frequency energy spectral density $\PP_\omega(x)$ 
for the spatiotemporal wave field $\psi(x,t)$ as
\begin{align}
\label{eq:psd}
\PP_k(t) = |a_k(t)|^2, \qquad \PP_\omega(x) = |a_\omega(x)|^2,
\end{align}
or alternatively we can add additional length or time normalizations
to convert those quantities to power spectral densities instead.

The first term $\Gamma a a^\dag$ in \cref{eq:H} denotes the
harmonic (quadratic) part of the Hamiltonian with either the complex
valued frequency $\Gamma=i\omega +\gamma$ or the wave number $\Gamma=i
k +\lambda$ that both include a pure oscillatory parts ($\omega$ or
$k$) and possible weakly excitation or damping rates, either temporal
$\gamma$ or spatial $\lambda$.  The second anharmonic term is cubic in
the lowest order of nonlinearity and describes the interactions
between various propagating and nonpropagating wave modes, where
$\alpha$, $\beta_a$ and $\beta_{a^\dag}$ are the complex valued
strengths of those different nonlinear processes.

An equation for the nonlinear oscillatory amplitude $a$ then can be
expressed as a derivative of the Hamiltonian form 
\begin{align}
\label{eq:a}
\der{a}{t}=\derp{H^s}{a^\dag}\equiv \Gamma a + \beta_{a^\dag} a
a^\dag + \beta_a a^2 - \alpha a (a a^\dag )^{1/2},
\end{align}
after removing the constants with a substitution of
$\beta_{a^\dag}=1/2 \tilde{\beta}_{a^\dag}$ and
$\alpha=1/3\tilde{\alpha}$ and dropping the tilde. We note
that although \cref{eq:a} is an equation for the temporal
evolution, the spatial evolution of the mode amplitudes $a_\omega(x)$
can be described by a similar equation substituting temporal
variables by their spatial counterparts, i.e., $(t,\omega,\gamma)
\rightarrow (x,k,\lambda)$.

Splitting \cref{eq:a} into an amplitude/phase pair of
equations using $a=Ae^{i\phi}$, assuming 
$\beta_a=\tilde{\beta_a}e^{-i\delta_a}$,
$\beta_{a^\dag}=\tilde{\beta}_{a^\dag} e^{i\delta_{a^\dag}}$,
and scaling the variables as 
\begin{align}
\label{eq:sc}
A=\gamma \tilde{A},\quad t=\frac{\tau}{\gamma}, \quad \omega=\tilde{\omega}\gamma,
\end{align}
gives the set of equations
\begin{align}
\label{eq:A0}
\der{\A}{\t} &=
\A + \A^2 \paren{\beta_{a^\dag}  \cos\Omega_{a^\dag}
+\beta_{a} \cos\Omega_a - \alpha } 
\\
\label{eq:B0}
\der{\phi}{\t} &=\o + \A \paren{-\beta_{a^\dag} \sin\Omega_{a^\dag}
+ \beta_{a} \sin\Omega_a} \
\end{align}
where $\Omega_a \equiv \phi-\delta_{a}$, $\Omega_{a^\dag} \equiv
\phi-\delta_{a^\dag}$.

These equations can further be cast into a more compact form as
\begin{align}
\label{eq:AC}
\der{\A}{\t} &=
\A + \A^2\left[R_a \cos{(\phi-\Phi)}- \alpha\right],
\\
\label{eq:BC}
\der{\phi}{\t} &=\o + \A R_\phi \cos{\phi},
\end{align}
where 
\begin{align}
R_a &= \sqrt{X_a^2+Y_a^2},\quad
R_\phi = \sqrt{X_\phi^2+Y_\phi^2},\\
\Phi_a &= \arctan{\frac{Y_a}{X_a}},\quad
\Phi_\phi = \arctan{\frac{Y_\phi}{X_\phi}},\\
\Phi& = \Phi_a-\Phi_\phi,
\end{align}
and
\begin{align}
X_a &=\phantom{-}\beta_{a^\dag}\cos{\delta_{a^\dag}} + \beta_{a}\cos{\delta_{a}},\n
Y_a &=\phantom{-}\beta_{a^\dag}\sin{\delta_{a^\dag}} + \beta_{a}\sin{\delta_{a}}, \n
X_\phi &=\phantom{-}\beta_{a^\dag}\sin{\delta_{a^\dag}} - \beta_{a}\sin{\delta_{a}},\n
Y_\phi &=-\beta_{a^\dag}\cos{\delta_{a^\dag}} + \beta_{a}\cos{\delta_{a}},\nonumber
\end{align}

An equilibrium 
(i.e.,~$d\A/d\t=d\phi/d\t=0$) 
solution of \cref{eq:AC,eq:BC} 
can be found from
\begin{align}
-\frac{R_\phi}{\o}\cos{\phi} +
R_a\cos{(\phi-\Phi)} - \alpha  
= 0,
\end{align}
as $\phi_e=\phi_0\equiv$ const and $\A_e=\o/R_\phi\cos{\phi_0}\equiv$
const.  This shows that for $\alpha>R_a$ there exist
critical values of $\o$ and $A_e$, such that
\begin{align}
\o_c&=\frac{R_\phi}{\alpha+R_a\cos{\Phi}},\quad \A_c =\o_c/R_\phi
\end{align}
which can also be expressed in terms of critical value of
one of the unscaled variables, either $\omega$ or $\gamma$
\begin{align}
\omega_c&=\gamma\o_c,
\quad \mathrm{or}\quad
\gamma_c=\frac{\omega}{\o_c},
\end{align}
This equilibrium solution provides the locus of the
bifurcation point at where the nonlinear spiking oscillations
occur (as
was shown both in \citep{Galinsky:2020a,*Galinsky:2020b} and in
\citep{Galinsky:2021a}).

The effective period of spiking $\mathcal{T}_s$ (or its inverse --
either the firing rate 1/$\mathcal{T}_s$ or the effective
firing frequency $2\pi/\mathcal{T}_s$) can be estimated from
\cref{eq:BC} by substituting $\A_c$ for $\A$ (assuming that the change
of amplitude $\A$ is slower than the change of the phase $\phi$) as
\begin{align}
\mathcal{T}_s &= \int\limits_0^{2\pi}\frac{d\phi}{\o + \o_c\cos{\phi}} = 
\frac{2\pi}{\sqrt{\o^2-\o_c^2}},
\end{align}
giving the unscaled effective spiking period $T_s$ and the
effective firing frequency $\omega_s$
\begin{align}
\label{eq:T}
T_s &= \frac{\mathcal{T}_s}{\gamma} =
\frac{2\pi}{\omega\sqrt{1-{\gamma^2}/{\gamma_c^2}}} 
= \frac{2\pi}{\omega\sqrt{1-\omega_c^2/\omega^2}},\\
\omega_s &= \frac{2\pi}{T_s} =\omega\sqrt{1-\omega_c^2/\omega^2},
\label{eq:o}
\end{align}
with the periodic amplitude $\A$ reaching the maximum
$\A_{max}=1/(\alpha-R_a)$ and the minimum $\A_{min}=1/(\alpha+R_a)$
for $d\A/d\t=0$ when $\phi=\Phi$ and $\phi=\Phi+\pi$ respectively.

\begin{figure}[!tbh] \centering
\includegraphics[width=1.0\columnwidth]{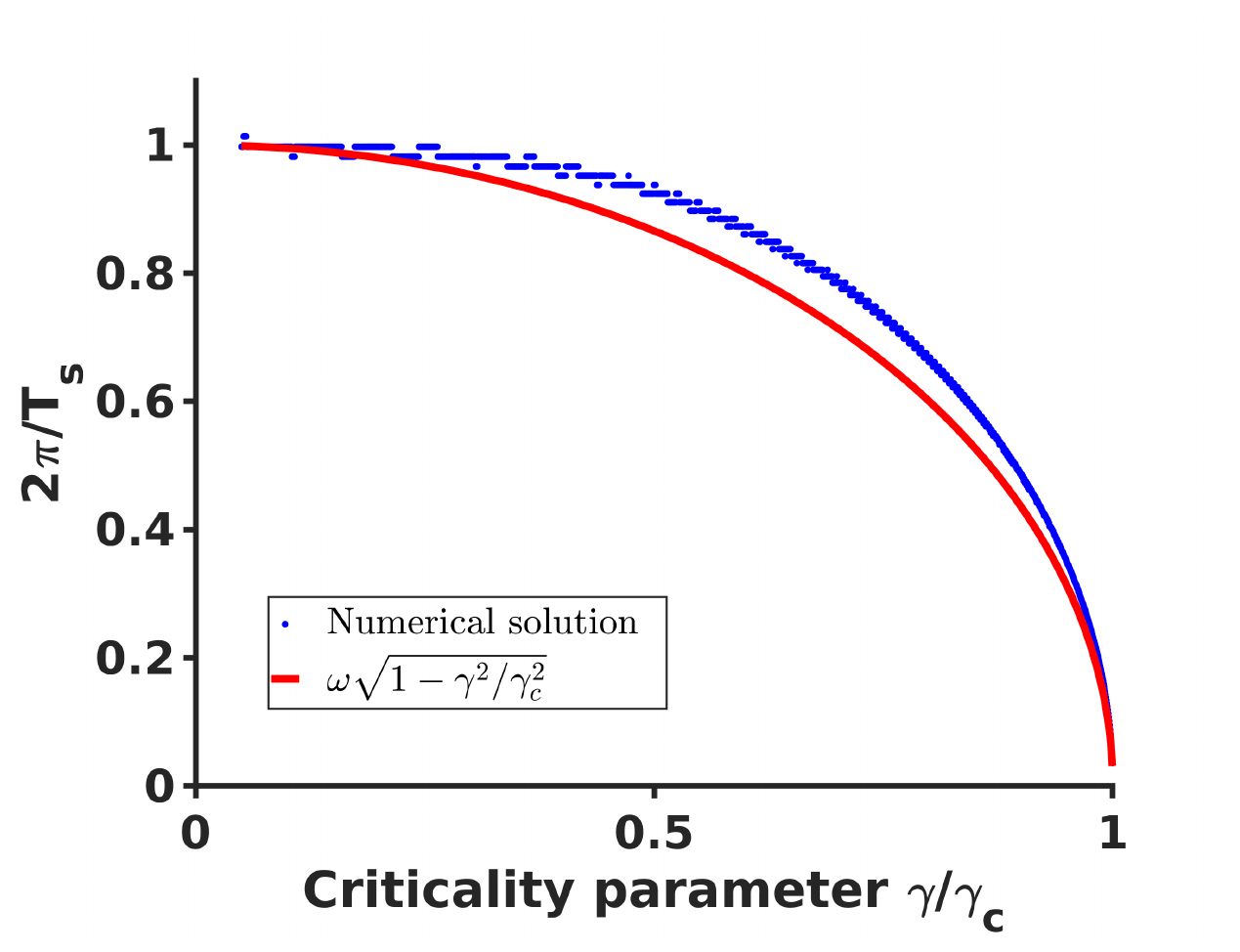}
\caption[]{Comparison of the analytical expression \cref{eq:o} for the
  effective spiking frequency $\omega_s=2\pi/T_s$ (red) and the
  frequency estimated from numerical solution of \cref{eq:AC,eq:BC}
  (blue) as a function of the criticality parameter $\gamma/\gamma_c$.
  In the numerical solution only $\gamma$ was varied and the
  remaining parameters were the same as parameters reported in
  \citep{Galinsky:2021a}.}
\label{fig:Period}
\end{figure}

The expressions \cref{eq:T,eq:o} are more general than typically used
expressions for the scaling exponent in the close vicinity
$|\gamma-\gamma_c|\ll \gamma_c$ of the critical point
\citep{kuramoto2013chemical,*pmid10057530,*pmid10058476}. They allow
recovery of the correct $T$ limits both at $\gamma \rightarrow
\gamma_c$ with the familiar $T\sim 1/\sqrt{\gamma_c-\gamma}$ scaling
and at $\gamma\sim0$ with the period $T$ approaching $T_0$ as $T\sim
T_0+O(\gamma^2) \equiv 2\pi/\omega+O(\gamma^2)$, where $T_0$ is the
period of linear wave oscillations with the frequency $\omega$.  In
the intermediate range $0<\gamma<\gamma_c$ the expressions
\cref{eq:T,eq:o} show reasonable agreement (\cref{fig:Period}) with
peak--to--peak period/frequency estimates from direct simulations of
the system \cref{eq:A0,eq:B0}.

Taking into account that the initial phase of spiking solutions of
\cref{eq:AC,eq:BC} is a random variable uniformly distributed on
$[0,2\pi]$ interval, the probability that a spike (either positive or
the more frequently experimentally reported negative
\citep{pmid14657176,*pmid15175392}) with duration width $\delta t_s$
and with 
the total period between the spikes ($T_s$)
will be detected is
simply $\delta t_s/T_s$ -- where the distance between spikes is
determined as the time interval needed for 2$\pi$ radian phase change,
that is the effective spiking period $T_s$.  Assuming
initially that the spike width $\delta t_s$ does not change when
approaching the critical point $\omega_c$, $\delta t_s$ can be
approximated by some fixed fraction of the linear wave period, i.e.,
$\delta t_s \sim \pi/\omega$, that gives for the probability density
\begin{align}
\label{eq:Pomega}
\P_k^{\{\omega\}}(\omega) \sim \omega^{-1}\sqrt{\omega^2/\omega_c^2-1},
\end{align}
for every wave mode with the wavenumber $k$.
It should be noted that the probability density $\P_k^{\{\omega\}}(\omega)$ 
has no relation and should not be confused with the frequency energy spectral density $\PP_\omega(x)$ (or with the power spectral density).

Transforming the frequency dependence of the wavenumber spectra $\P_k^{\{\omega\}}(\omega)$ to the temporal
domain ($T=2\pi/\omega$, $T_c=2\pi/\omega_c$)
\begin{align}
\int\limits_{\omega_c}^\infty \P_k^{\{\omega\}}(\omega)d\omega = 
\int\limits_0^{T_c}\P_k^{\{\omega\}}\left(\frac{2\pi}{T}\right)\frac{2\pi}{T^2} dT =
\int\limits_0^{T_c}\P_k^{\{T\}}(T) dT,
\end{align}
gives for the temporal probability density $\P_k^{\{T\}}(T)$
\begin{align}
\label{eq:PT}
\P_k^{\{T\}}(T) \sim T^{-2}\sqrt{1-T^2/T_c^2},
\end{align}
hence the scaling of the temporal probability density $\P_k^{\{T\}}$ follows the power
law with -2 exponent with additional $\sqrt{1-T/T_c}$ falloff in close
vicinity of the critical point in agreement with temporal scaling of
neuronal avalanches reported in \citep{pmid14657176,*pmid15175392}.

The above single wave mode analysis shows that the probability density
$\P_k^{\{T\}}$ for any arbitrary selected wave mode $k$ with arbitrary
chosen threshold follows power law distribution with -2 exponent,
therefore, a mixture of multiple wave modes that enters into the
spatiotemporal wave field $\psi(x,t)$ with different amplitudes and
different thresholds will again show nothing more than the same power
law distribution.

Due to the reciprocity of the temporal and spatial representations of
the Hamiltonian form \cref{eq:H} equations for the spatial wave
amplitude have the same form as the temporal equations
\cref{eq:AC,eq:BC}
\begin{align}
\label{eq:ACS}
\der{\A}{\xi} &=
\A + \A^2\left[R_a \cos{(\phi-\Phi)}- \alpha\right],
\\
\label{eq:BCS}
\der{\phi}{\xi} &=\k + \A R_\phi \cos{\phi}, 
\end{align} 
under similar scaling (the spatial equivalent of
\cref{eq:sc}) of the wave amplitude, the coordinate, and the wave
number
\begin{align}
\label{eq:scS}
A=\lambda \tilde{A},\quad x=\frac{\xi}{\lambda}, \quad k=\tilde{k}\lambda.
\end{align}
In the spatial domain, this leads to the critical parameters
$\A_{c}$ and $\k_{c}$
\begin{align}
\k_c&=\frac{R_\phi}{\alpha+R_a\cos{\Phi}},\quad \A_c =\k_c/R_\phi.
\end{align}
Although our
simple one dimensional scaling estimates do not take into account the
intrinsic spatial scales of the brain, e.g., cortex radius of
curvature, cortical thickness, etc., nevertheless, even in this
simplified form the similarity between spatial and temporal nonlinear
equations suggests that the nonlinear spatial wave behavior will
generally look like spiking in the spatial domain where some localized
regions of activity are separated by areas that are relatively signal
free and this separation will increase near the critical
point. Exactly this behavior was reported in the original experimental
studies of the neuronal avalanches \citep{pmid14657176,*pmid15175392},
where it was stated that the analysis of the contiguity index revealed
that activity detected at one electrode is most often skipped over the
nearest neighbors, but this experimental observation of near critical
nonlinear waves was instead presented as the indicator that the
activity propagation is not wave-like. The effects of the
intrinsic spatial scales of the brain will certainly affect the
details of the spatial critical wave dynamics and so their inclusion
will be important for more completely characterizing the details of
brain criticality and will be the focus of future investigations.

Using the spatial equations \cref{eq:ACS,eq:BCS} similar scaling
results can be obtained for the wave number $k$ and the linear spatial
dimension $L$ probabilities for every wave mode with the frequency $\omega$ as
\begin{align}
\label{eq:Pk}
\P_\omega^{\{k\}}(k) &\sim k^{-1}\sqrt{k^2/k_c^2-1},\\
\label{eq:PL}
\P_\omega^{\{L\}}(L) &\sim L^{-2}\sqrt{1-L^2/L_c^2},
\end{align}
where $L$ is the linear spatial scale related to the wave number as $k=2\pi/L$.

The linear spatial dimension of the avalanche $L$ is related to its
area $S$ on a 2 dimensional surface as $L=\sqrt{S}$, hence
\begin{align}
\int\limits_{0}^{L_c} \P_\omega^{\{L\}}(L)dL = 
\int\limits_0^{S_c}\frac{\P_\omega^{\{L\}}\left(\sqrt{S}\right)}{2\sqrt{S}} dS =
\int\limits_0^{S_c}\P_\omega^{\{S\}}(S) dS,
\end{align}
\begin{align}
\label{eq:PS}
\P_\omega^{\{S\}}(S) \sim S^{-3/2}\sqrt{1-S/S_c},
\end{align}
hence the spatial probability scaling for the size $S$ follows the
power law with -3/2 exponent again with additional $\sqrt{1-S/S_c}$
falloff in close vicinity of the critical point, that is also in
agreement with experimentally reported spatial scaling of neuronal
avalanches \citep{pmid14657176,*pmid15175392}.  We would like to
mention that the nonlinear anharmonic oscillations described by the
\cref{eq:AC,eq:BC} only exists for frequencies and wave numbers that
are above the critical frequency $\omega_c$ or the critical wave
number $k_c$ values that define maximal possible temporal $T_c$ or
spatial $L_c$ scales of the nonlinear oscillations. If the finite
system sizes are below those maximal values the cutoffs will be
defined by the system scales.

The assumption of the fixed spike duration $\delta t_s$ used in
\cref{eq:Pomega,eq:PT} (or the spike length for spatial spiking in
\cref{eq:Pk,eq:PL}) can be improved by estimating the scaling of the
spike width as a function of the criticality parameter from the
amplitude equation (we will use the temporal form of the equation
\cref{eq:AC} but the spatial analysis with equation \cref{eq:ACS} is
exactly the same).

Dividing equation \cref{eq:AC} by $\A$ and taking an integral around some area 
in the vicinity of the amplitude peak $\A_{max}$ we can write
\begin{align}
\int\limits_{\A_-}^{\A_+}\frac{1}{\A}d\A = 
\int\limits_{\t_-}^{\t_+}d\t +
\int\limits_{\Phi_-}^{\Phi_+}\frac{\o_c}{R_\phi}
\frac{R_a \cos{(\phi-\Phi)}-\alpha}{\o+\o_c\cos{\phi}} d\phi,
\end{align}
where $\t_{\pm} = \t_{max} \pm \delta\t$, and $\t_{max}$ is the
location of spiking peak. Neglecting the spike shape asymmetries,
i.e., assuming that $\t_{\pm}$ correspond to symmetric changes in both
the amplitudes $\A_{\pm}=\A(\t_\pm)=\A_{max}-\delta\A$, and the phases
$\Phi_{\pm} = \Phi(\t_\pm) = \Phi\pm\delta\Phi$, we can then estimate
the spike duration $\delta t_s\equiv (\t_+-\t_-)/\gamma$ as
\begin{align}
\delta t_s = \frac{1}{\gamma} 
\int\limits_{\Phi-\delta\Phi}^{\Phi+\delta\Phi}
\frac{1-R(\cos(\Phi)+\cos{(\phi-\Phi))}}{\o+\o_c\cos{\phi}} d\phi,
\label{eq:ts}
\end{align}
where, similar to the spiking period estimation in \cref{eq:T}, we
again assume that the main contribution
comes from the change of the oscillation phase, hence
$\A_c$ can be substituted for $\A$.  For $\delta\Phi$ some fixed value
that is smaller or around a quarter of the period (i.e.,
$\delta\Phi\lesssim\pi/2$) can be chosen, and $R=\o_c R_a/R_\phi$.

\begin{figure}[!tbh] \centering
\includegraphics[width=1.0\columnwidth]{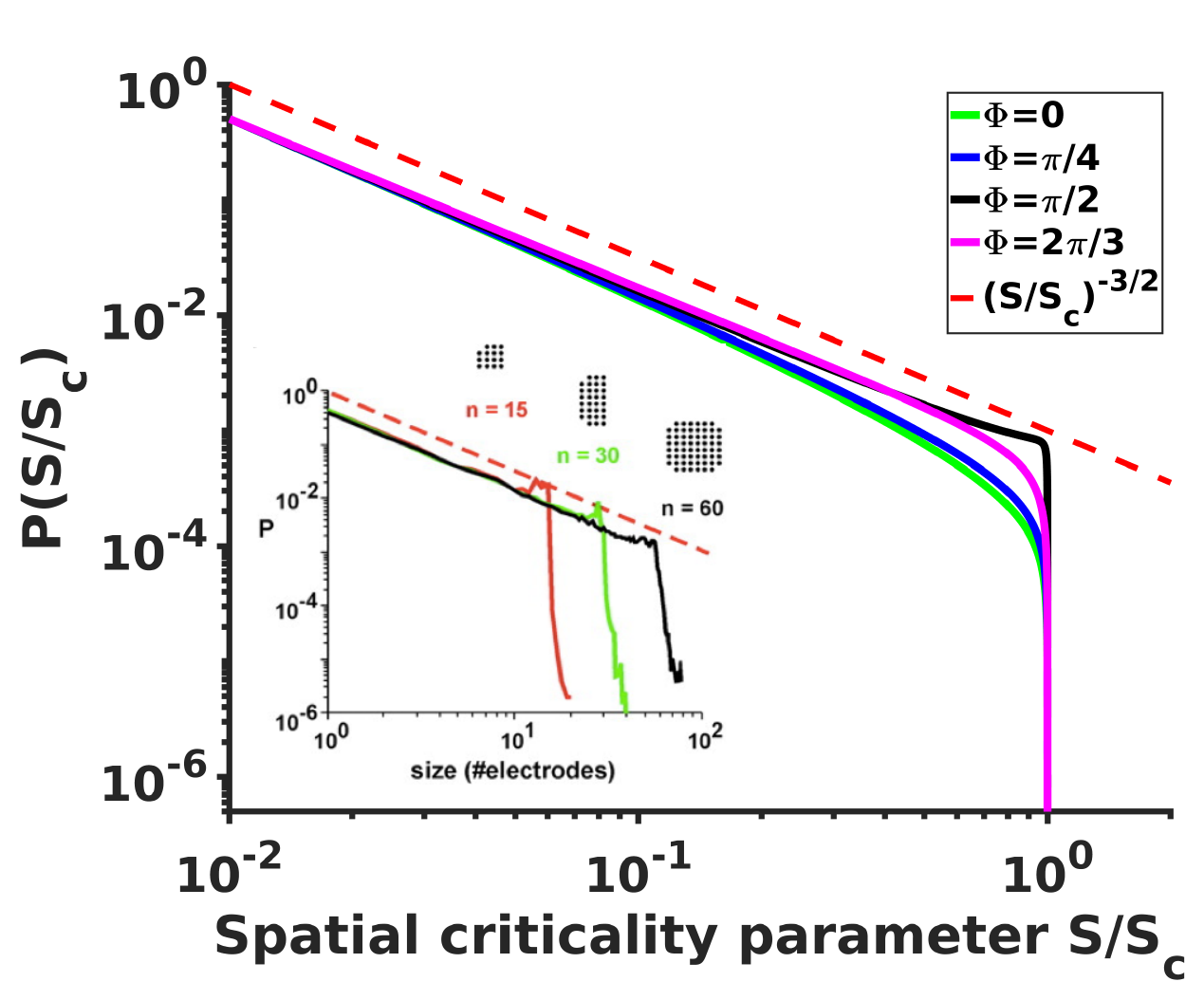}
\caption[]{Analytical probability density spectra as a function of brain waves
  criticality parameter $S/S_c$ show excellent agreement with the
  experimental avalanche data (insert, from
  \citep{pmid14657176,*pmid15175392}) reproducing not only the overall
  shape of the spectra with the -3/2 power exponent at the initial
  scale free part of the spectra
  and the steep falling edge in the vicinity of the critical
  point, but also reproduce the fine details such as the small
  bump-like flattening of the spectra at the transition from -3/2 leg
  to the steep falling edge that is clearly evident in experimental
  spectra.  }
\label{fig:S}
\end{figure}

An expression \cref{eq:ts} can be evaluated in closed form but we do
not include it here and instead plotted the final spatial probability density
spectra $\P(S/S_c)$, similarly obtained from the expression for $\delta l_s/L_s$
again substituting $L=\sqrt{S}$ and $dL=dS/(2\sqrt{S})$,
for several values of the phase shift $\Phi$ (\cref{fig:S}). The
spectra clearly show again the same power law dependence with -3/2
exponent as was reported in \citep{pmid14657176,*pmid15175392}
followed by a steep falloff sufficiently close to the critical
point. What is interesting, however, is that the spectra for
$\Phi=\pi/2$ (and this is the phase shift value used for spiking
solutions reported in
\citep{Galinsky:2020a,*Galinsky:2020b,Galinsky:2021a}) recover even
the fine structure of the scaling and clearly show the small bump at
the end of the scale free part of the spectra where the local probability
deflects from the initial -3/2 power exponent and flattens first
before turning in to the steep falloff. These small bumps are evident
in all experimental spectra \citep{pmid14657176,*pmid15175392} shown
on the insert in \cref{fig:S}.

In summary, in this Letter we have presented an analysis of
temporal and spatial probability density spectra that are generated due to the
critical dynamics of the nonlinear weakly evanescent cortical wave
(WETCOW) modes \citep{Galinsky:2020a,*Galinsky:2020b}.  The
Hamiltonian framework developed for these WETCOW modes in
\citep{Galinsky:2021a} is advantageous in that it explicitly
uncovers the reciprocity of the temporal and the spatial
dynamics of the evolutionary equations. Therefore, in the nonlinear regime
sufficiently close to the critical point the spatial behavior of the
wave modes displays features similar to the properties of their
nonlinear temporal dynamics that can be described as spatial domain
spiking, with localized regions of wave activity separated by
quiescent areas, with this spatial spiking intermittence
increasing near the critical point. Similar spatial
behavior was observed experimentally in neuronal avalanches, when
activity detected at one electrode was typically skipped over the
nearest neighbors.  This was interpreted as evidence that
avalanche spatial intermittency is not wave-like in
nature \citep{pmid14657176,*pmid15175392}.
Our results demonstrate the contrary,
however: the spatial patterns are the direct result of nonlinear
interactions of weakly evanescent cortical waves.

Both temporal and spatial scaling expressions analytically estimated
from the nonlinear amplitude/phase evolutionary equations show
excellent agreement with the experimental neuronal avalanche probability
spectra reproducing not only the general average power law exponent
values and falloffs in the vicinity of the critical point, but also
finding some very subtle but nevertheless clearly experimentally
evident fine details, like bumps in the transition region at the edge
of the scale free part of the probability spectra.

The brain wave model thus uncovers the physical processes
behind the emergence of avalanches that were hidden for almost 20
years since their discovery and demonstrates that the power scaling
property of the neuronal avalanches can be explained by the same
criticality that is responsible for spiking events in either space or
time and is a consequence of a nonlinear interaction of weakly
evanescent transverse cortical waves (WETCOWs,
\citep{Galinsky:2020a,*Galinsky:2020b}).  
The origin of these 'avalanche' properties emerges
directly from the same theory that produces the wide range of oscillatory,
synchronized, and wave-like network states, and does not require one
to posit the existence of either new brain
activity states, nor construct analogies between brain activity and
{\it ad hoc} generic 'sandpile' models \citep{Galinsky:2021Supp}.

In a more general way these results may be applicable not only to neuronal avalanches
but to many other physical systems that involve wave processes as they 
show that a system of wave modes interacting through all
possible combinations of the third order nonlinear terms described by
a general wave Hamiltonian necessarily produces anharmonic wave modes
with temporal and spatial scaling properties that follow scale free
power laws.

\begin{acknowledgments}
LRF and VLG were supported by NSF grant ACI-1550405, UCOP MRPI grant MRP17454755
and NIH grant R01 AG054049.
\end{acknowledgments}


%

\end{document}